\documentclass[aps,prb,superscriptaddress,twocolumn]{revtex4}
\usepackage{amsmath}
\usepackage{amssymb}
\usepackage{amsthm}
\usepackage{graphicx}
\usepackage{graphics}
\usepackage{subfigure}
\usepackage{bm}
\usepackage{hyperref}
\usepackage{rotating}
\usepackage{color}

\def\beas{\begin{eqnarray*}}
\def\eeas{\end{eqnarray*}}
\def\bea{\begin{eqnarray}}
\def\eea{\end{eqnarray}}
\def\be{\begin{equation}}
\def\ee{\end{equation}}

\newcommand{\ua}{\uparrow}
\newcommand{\da}{\downarrow}
\newcommand{\down}{\downarrow}

\begin{document}
\centerline{   {\bf }}

\title{Interference, Coulomb blockade, and the identification of non-abelian quantum Hall states }

\author{Ady Stern}
\affiliation{Department of Condensed Matter Physics, Weizmann
Institute of Science, Rehovot 76100, Israel}
\author{Bernd Rosenow}
\affiliation{Max-Planck-Institute for Solid State Research,
Heisenbergstr. 1, D-70569 Stuttgart, Germany }
\author{Roni Ilan}
\affiliation{Department of Condensed Matter Physics, Weizmann
Institute of Science, Rehovot 76100, Israel}
\author{Bertrand I. Halperin }
\affiliation{ Physics Department, Harvard University, Cambridge
02138, Massachusetts, USA}

\date{\today}
\begin{abstract}

We study electronic transport phenomena in a Fabry-Perot
interferometer in the fractional quantum Hall regime in two
limits. We analyze the lowest-order interference pattern in a
nearly open interferometer (weak-backscattering limit)  and the
temperature-dependence of the Coulomb-blockade transmission peaks
in a nearly closed interferometer (strong-backscattering limit).
For both limits we consider two series of fractional quantized
Hall states, one with abelian and one with non-abelian
quasiparticles. We show that the results obtained in the two
limits give identical information about the quasiparticle
statistics. Although the experimental signatures of the abelian
and non-abelian states may be similar in some circumstances, we
argue that the two cases may be distinguished due to the
sensitivity of the abelian states to local perturbations, to which
the non-abelian states are insensitive.

\end{abstract}

%\pacs{}

\maketitle

\section{Introduction}

Much effort has been devoted in recent years to the search for
experimental demonstrations of the exotic quantum statistics of
quasi-particles in the fractional quantum Hall effect. Experiments
were proposed, and some have been carried out, or attempted. Among
these there are several that are based on the quantum Hall analog
of the Fabry-Perot interferometer, either through
interference\cite{Fradkin,SternHalperin,Bonderson-5/2,Bonderson-RR,stone,
ilan-2008} or through the Coulomb
blockade\cite{SternHalperin,ilan-2007,ilan-2008}. In particular,
interference and Coulomb blockade were predicted to show rather
robust signatures of non-abelian statistics in quantum Hall states
that are believed to be non-abelian. Recent data on a Fabry-Perot
interferometer at the $\nu=5/2$ state may be a first confirmation
of these predictions\cite{willett-2008}.

In the context of the quantum Hall effect (QHE), a Fabry-Perot
interferometer, shown schematically in Fig. (1), is a Hall bar
perturbed by two constrictions (quantum point contacts -
QPCs)\cite{Chamon}. The two QPCs introduce amplitudes for
inter-edge tunneling of quasi-particles. The quantity of interest
is the probability of back-scattering, as a function of magnetic
field and the interferometer's area. When the amplitudes for
inter-edge tunneling are small, the probability for
back-scattering involves the interference of two trajectories.
When the amplitudes are large, the interferometer is almost
closed, and its interior becomes a quantum dot. The probability
for back-scattering is then close to unity, except in the vicinity
of Coulomb blockade transmission peaks. The transition between the
two limits - the "lowest order interference" and "Coulomb
blockade" limits - may be tuned by adjusting the tunneling
amplitudes of the two point contacts.

\begin{figure}[t]
\begin{center}
\subfigure{\label{fig:inter-a}\includegraphics[width=0.4\linewidth,angle=-90]{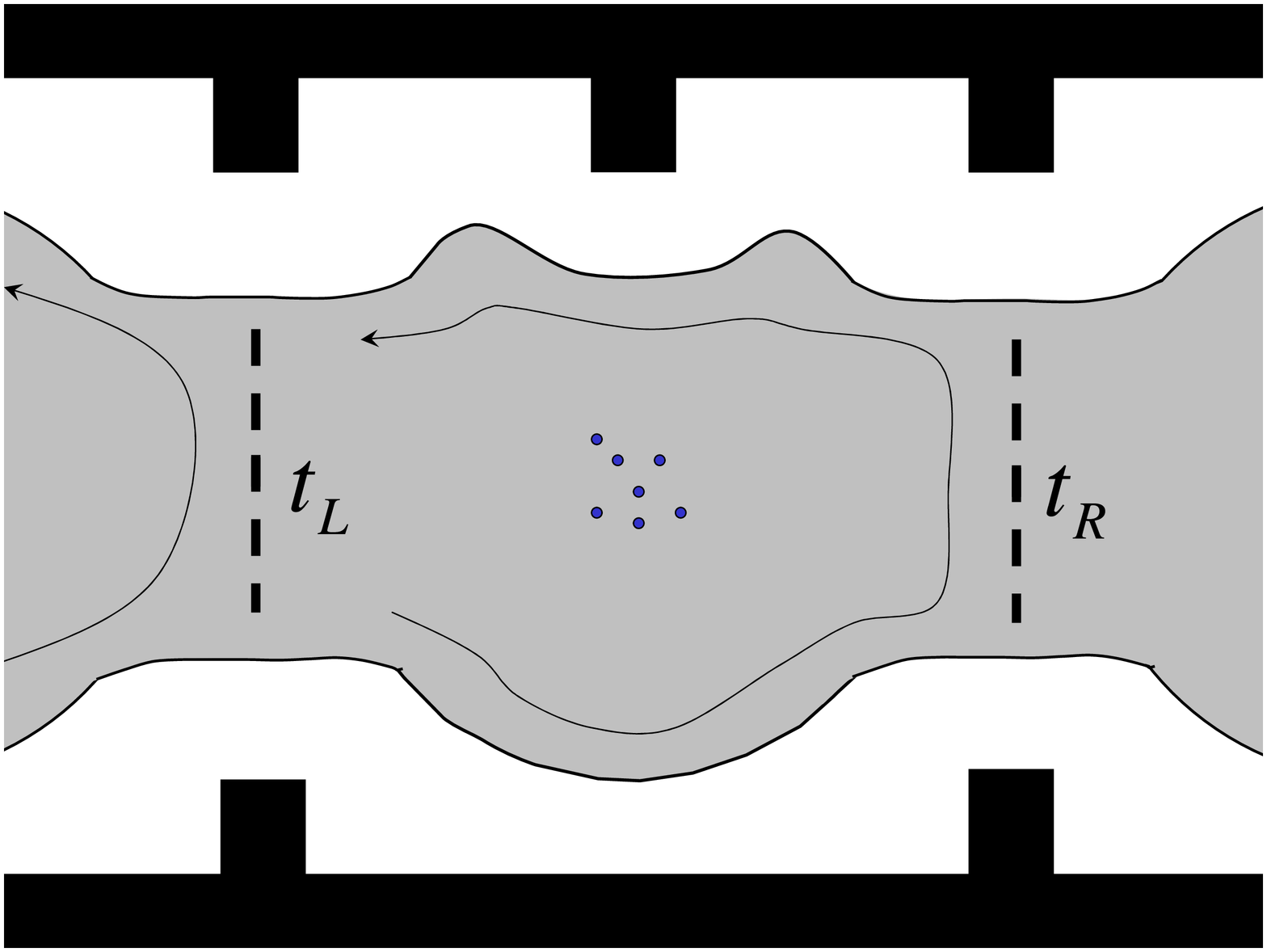}}\vspace{1cm}
\subfigure{\label{fig:inter-b}\includegraphics[width=0.4\linewidth,angle=-90]{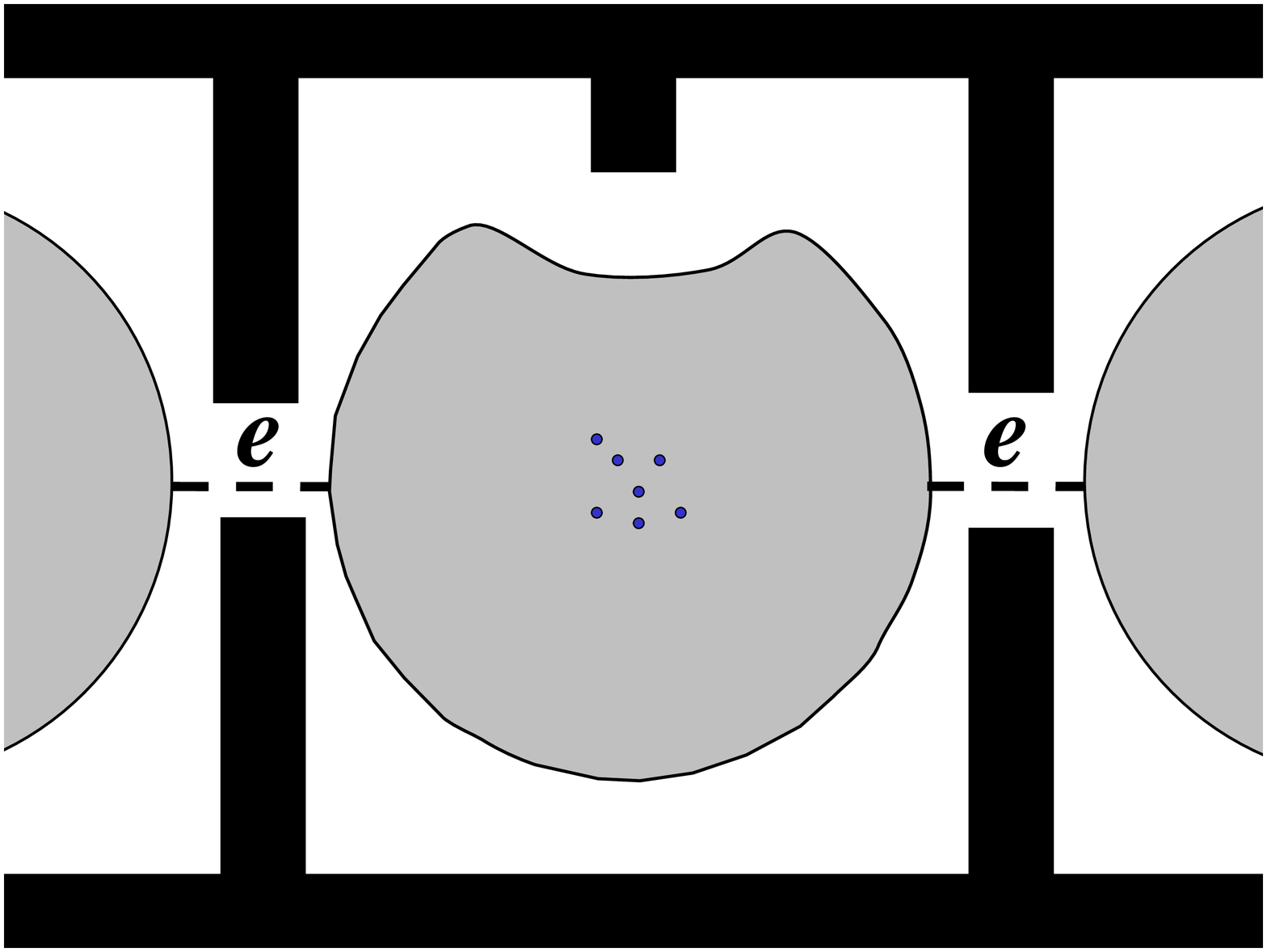}}
\end{center}\
\vspace{0.5cm}\caption{A sketch of the Fabry-Perot interferometer.
The upper figure shows the interferometer in the lowest order
interference limit. Quasi-particles tunnel from one edge to the
other at the two QPC's and the two interfering trajectories appear
above. The lower figure shows the interferometer in the Coulomb
blockade regime, where the QPC's are closed and only electrons are
allowed to tunnel through the dot. } \label{fig:interferometer}
\end{figure}

The expected dependence of the back-scattered current $I$ on the
magnetic field $B$ and the interferometer's area $A$ for
non-abelian QHE states was calculated in several
works\cite{SternHalperin,Bonderson-5/2,Bonderson-RR,stone,
ilan-2008}. Unique signatures were predicted, that originate from
the non-abelian nature of the states. These signatures all emerged
from a model in which the  bulk of the interferometer houses a
number of localized quasi-particles, $n_{is}$. The anyonic
statistics associated with the edge current that encircles these
quasi-particles modifies the interference contribution to the
back-scattered current. The model assumes that the area of the
interferometer may be varied by means of a voltage applied to a
gate, and the number $n_{is}$ may be varied by means of a magnetic
field. Furthermore, the model assumes that the density of
quasi-particles is sufficiently low such that the $n_{is}$
quasi-particles localized in the bulk are far enough from the edge
for their coupling to the edge to be negligible. The area of the
interferometer is assumed not to vary with the variation of a
magnetic field. Finally, the model assumes that the back-scattered
current primarily originates from tunneling of quasi-particles
whose charge is the smallest one possible, and whose tunneling is
most relevant (in the renormalization group sense). The
applicability of these assumptions to the systems used in current
experiments and the effect of deviations from these assumption are
a subject of current experimental\cite{YiMing2009,Ofek2009} and
theoretical\cite{RosenowPRL, RosenowPRB, BisharaPRB} studies. For
the present study, we adopt this model.

In this work we first study  $I(B,A)$ in the limits of lowest
order interference and of Coulomb blockade. We find that the two
limits give the same information regarding the state they probe.
We then get an insight to the relation between the two limits by
focusing on a Coulomb blockaded interferometer and studying the
thermally averaged number of electrons in the interferometer
${\cal N}(B,A,T)$ as a function of the field $B$, the area $A$ and
the temperature $T$. At low temperature, $\cal N$ is an integer
that rises in steps when the area is increased, and the derivative
$\partial{\cal N}/\partial A$ shows a series of peaks. The maxima
in the derivative  coincide with the Coulomb blockade peaks in the
conductance through the dot. As the temperature rises, these peaks
are smeared into a sinusoidal pattern. We find the dependence of
this sinusoidal pattern on the properties of the quantum Hall
state to be identical to that found in $I(B,A)$ in the lowest
order interference limit.

Generally, the discrete spectrum of a quantum dot is a result of a
Bohr-Sommerfeld interference of infinitely-many trajectories. At
low temperature, a small number of energy states is probed and
many trajectories interfere. As the temperature gets high,
interference of trajectories that encircle the dot many times is
smeared, until eventually only lowest order trajectories are left.
The unique properties of the quantum Hall states that we examine,
both abelian and non-abelian, allow us to relate the two limits.
Remarkably, although the energy states of the closed dots are all
characterized by an integer number of electrons, the high
temperature behavior of ${\cal N}(B,A)$ reflects the properties of
the quasi-particle with lowest charge.

Following a recent work by Bonderson et al. \cite{bonderson-2009}
we also examine  the level of unambiguity with which the
Fabry-Perot interferometer is able to identify a non-abelian
quantum Hall state, namely, we examine whether  the same
experimental signatures may result from states that are abelian as
well as non-abelian. Bonderson and collaborators have shown that
the zero temperature Coulomb blockade peak patterns that are
predicted for the non-abelian Read-Rezayi series of states are
identical to the ones predicted for the abelian series of
multi-component Halperin states\cite{bonderson-2009}. We show that
for the most prominent candidate for a non-abelian state,
$\nu=5/2$, the same identity of patterns holds also for
Fabry-Perot interference experiments and for finite temperature
Coulomb blockade experiments. Largely, this holds also for the
more complicated states at $\nu=2+\frac{k}{k+2}$ with $k>2$,
although in this case subtle differences exist in the interference
signals between the abelian multi-component Halperin states and
the non-abelian Read-Rezayi states.

The identity of the predictions for these two series of abelian
and non-abelian states is based on a very limiting assumption
regarding the abelian states. The abelian multi-compnent states of
$\nu=2+\frac{k}{k+2}$ are composed of $k$ flavors of electrons,
and their similarity to the non-abelian Read-Rezayi states holds
only under a full symmetry between all the flavors. A breaking of
that symmetry affects the Coulomb blockade and interference
patterns and distinguishes the abelian and non-abelian
states\cite{bonderson-2009}. In the $k=2$ case, corresponding to
the $\nu=5/2$ state, the two flavors of electrons are likely to be
the two spin states. The symmetry between the two states would
then be broken by the Zeeman coupling of the spin to the magnetic
field, or by spin-orbit coupling. In contrast, the patterns
predicted for the non-abelian states enjoy the insensitivity
characteristic of these states to local perturbations.  For the
$k>3$ cases, there are no obvious degrees of freedom that lead to
the electronic system splitting into $k$ flavors, but we find the
comparative analysis of the multi-component Halperin states and
the Read-Rezayi states to be of theoretical interest. Again,
predictions for the abelian and non-abelian series of states are
very similar at the point of exact symmetry between the $k$
flavors, and differ as this symmetry is broken. Again, the abelian
states are sensitive to local perturbations, while the non-abelian
ones are not. We note that both abelian and non-abelian states are
sensitive to a coupling between the localized bulk quasi-particles
and the chiral neutral modes, as discussed in
\onlinecite{BisharaPRB,RosenowPRB,RosenowPRL}.

The structure of the paper is as follows: in Sec. II we analyze
the $\nu=5/2$ case. We examine two candidate states for which the
Coulomb blockade signatures were found to be identical, the
Pfaffian and the Halperin $(331)$ state. We show that the same
holds for the interference pattern as well, and comment on the
subtle differences between the two states, differences that hold
for both the Coulomb blockade and the interference experiments. In
Sec. III we analyze the $\nu=2+\frac{k}{k+2}$ series. We compare
the Read-Rezayi series\cite{ReadRezayi} to the multi-component
Halperin state series\cite{WenZee}, and show that the similarity
in Coulomb blockade patterns, discovered in
[\onlinecite{bonderson-2009}], largely holds also for the FP
interference. In Sec IV we study  a Coulomb blockaded quantum dot
and ask how the thermally averaged number of electrons in the dot
${\cal N}(B,A)$ depends on the field $B$, the area $A$ and the
temperature $T$.

\section{Interference signals in the $\nu=5/2$ state}

The two leading candidate states for a non-abelian $\nu=5/2$
phase, the Pfaffian\cite{MooreRead} and the
anti-Pfaffian\cite{AntiPf1,AntiPf2}, have been predicted to show a
unique behavior in an FP device, in both limits. In the limit of
lowest order interference\cite{SternHalperin,Bonderson-5/2} the
pattern depends crucially on the parity of $n_{is}$. For an odd
$n_{is}$, no interference signal is to be seen, i.e., the
back-scattered current would show no periodic dependence on the
area of the interferometer. For an even $n_{is}$ a periodic
dependence should be observed, with the phase of the interference
pattern assuming one of two possible values, mutually shifted by
$\pi$. The phase chosen depends on the topological charge of the
$n_{is}$ localized quasi-particles. In the Coulomb blockade
limit\cite{SternHalperin} the area spacing between two consecutive
Coulomb blockade peaks depends on the parity of $n_{is}$. For odd
$n_{is}$ the peaks are equally spaced, while for even $n_{is}$
they bunch into pairs.

These predictions are rather unique, being different from those
expected for states of the IQHE and simple FQHE
states\cite{Cappelli-2,Cappelli-1,Chamon}. However, they alone do
not identify the $\nu=5/2$ state as non-abelian, since another
candidate state, the $(3,3,1)$ state\cite{331}, shares the same
features. It was already shown that the Coulomb blockade patterns
of the Pfaffian, Anti-Pfaffian and the $(3,3,1)$ are
identical\cite{bonderson-2009}. We now show that the same holds
for the lowest order interference.

The $(3,3,1)$ state is a paired state characterized by a
$K$-matrix\cite{wen} of the form \bea K=\left(
\begin{array}{cc}
3 & 1 \\
1 & 3 \\
\end{array}
\right) \eea with the quasi-particle operators characterized by
the vectors $l_\ua=(1,0)$ and $l_\down=(0,1)$. The $n_{is}$
quasi-particles are characterized by the vector
$n=(n_\ua,n_\down)$, with $n_{is}=n_\ua+n_\down$. When an $l_i$
(with $i=\ua,\down$) quasi-particle encircles the bulk
quasi-particles it accumulates a phase of $2\pi l_i K^{-1}n.$ The
incoming current is spin-unpolarized, and thus the observed
interference pattern is the sum of two patterns, whose phase
difference is
\begin{equation}
2\pi(l_\ua-l_\down)K^{-1}n=\pi(n_\ua-n_\da) \label{phasediff}
\end{equation}
When $n_{is}$ is odd this phase difference is an odd multiple of
$\pi$. The two patterns then mutually cancel, leading to no
periodic dependence on area. In contrast, for $n_{is}$ even, the
two patterns interfere constructively, and an interference pattern
is to be seen. Furthermore, for an even $n_{is}$ either
$n_{\ua},n_\down$ are both even or both are odd. Interestingly,
the interference patterns that result in these two cases are
mutually shifted by $\pi$. These characteristics of the
interference patterns are the same as those of the non-abelian
states - vanishing interference for an odd $n_{is}$, and two
possible interference patterns, mutually shifted by $\pi$, for
even $n_{is}$.

Just as in the case of the Coulomb blockade, the interference
patterns described above for the $(3,3,1)$ crucially depend on the
symmetry between up and down spins. Any deviation from this
symmetry, for example in having a polarized incoming current or a
Zeeman splitting between the two types of localized
quasi-particles, would affect both the Coulomb blockade peaks and
the interference patterns. This is in an important contrast to the
non-abelian Pfaffian state, in which the unique Fabry-Perot
signatures are robust. For example, the vanishing lowest order
interference in the case of odd $n_{is}$ results in the $(3,3,1)$
case from the addition of two interference patterns, corresponding
to the two spin directions. If the two are not of equal weight,
they do not sum to zero. In contrast, in the Pfaffian case, the
vanishing of the interference for odd $n_{is}$ results from the
fusion of two $\sigma$ operators in the Ising conformal field
theory (CFT) to equal weights of $1$ and $\psi$ particles, whose
interference patterns are mutually shifted by
$\pi$.\cite{Bonderson-5/2} The equal weight of the two patterns is
in this case inherent to the description of the state by the Ising
CFT.

\section{Interference signals for the $\nu=2+\frac{k}{k+2}$ states}

Intriguing candidate description for filling factors
$\nu=2+\frac{k}{k+2}$ are the Read-Rezayi states\cite{ReadRezayi}.
These states have been predicted to have their own unique
characteristics in F-P
experiments\cite{ilan-2007,ilan-2008,Bonderson-RR}. In the limit
of lowest order interference, the observed interference pattern
depends on the topological charge $l$ of the $n_{is}$
quasi-particles localized in the bulk. For each value of $n_{is}$
there are several possible integer values of the topological
charge. For odd $k$, there are $(k+1)/2$ possible values of $l$.
For even $k$, there are either $k/2$ or $\frac{k}{2}+1$ possible
values depending on whether $n_{is}$ is odd or even. The amplitude
$I(l)$ of the interference term depends on $l$, being
%
%*********************  RR interference reduction **************
\begin{equation}
I(l)=\frac{\cos{\frac{\pi(l+1)}{k+2}}}{\cos\frac{\pi}{k+2}}
\label{visibilityrr}
\end{equation}
%***************************************************
%
As for the Coulomb blockade limit\cite{ilan-2007}, the positions
of the Coulomb blockade peaks as a function of the interferometer
area depend on the topological charge, as different patterns of
bunching of peaks are observed for different values of $l$.

Again, the Coulomb blockade peak spacings predicted for the
Read-Rezayi states are identical to those predicted for the
multi-component Halperin states. In view of the analogy described
above between the Coulomb blockade and lowest order interference
for the $k=2$ case, it is natural to examine the lowest order
interference pattern expected for the multi-component Halperin
state for a particular $k$. The analysis starts from the
$K$-matrix, which is now a $k\times k$ matrix, whose elements
satisfy $K_{ij}=1+2\delta_{ij}$, in a basis in which the charge
vector $t$ satisfies $t_i=1$ for all $i=1..k$. The state of the
bulk is now described by a vector $n=(n_1..n_k)$ with
$\sum_{i=1}^k n_i=n_{is}$. The most relevant quasi-particles are
described by the vectors $l^{(j)}$ ($j=1..k$), with the elements
of the vector $l^{(j)}$ being all zero, except the $j$'th element,
which is one: $l^{(j)}=(0..0, 1, 0..0)$. The phase accumulated by
a quasi-particle $l^{(j)}$ encircling the bulk is
%
%**************  interference phase from K-matrix **************
\be 2\pi l^{(j)}K^{-1}n=-\pi\frac{n_{is}}{(k+2)}+\pi n_j
\label{statisticsk}
\end{equation}
%***************************************************
%
As seen in this expression, the phases accumulated by different
types of quasi-particles are identical, {\it up to a possible
shift of $\pi$}. For a quasi-particle of type $l^{(j)}$ this shift
is present for odd $n_j$ and absent for even $n_j$. Again, if the
incoming current does not break the symmetry between the $k$ types
of electrons, then the observed interference pattern will be a sum
of $k$ patterns, some of which are $\pi$-shifted with respect to
the rest. If the vector $n$ is made of ${N_e}$ even numbers
and $N_o$ odd numbers (with $N_e+N_o=k$ and both non-negative),
and if we normalize the amplitude of the combined interference
pattern to be $1$ for the case $N_e=k$, then the amplitude of the
interference pattern for the case where the bulk is described by a
vector $n$ is
%
%**************  multi-component Halperin reduction ************
\begin{equation}
I(n)=\frac{k-2N_o}{k} \label{visibilityhalk}
\end{equation}
%************************************************************
%
The number of possible values for $I(n)$ depends on the parity of
$k$ and the parity of the number of bulk quasi-particles $n_{is}$.
In principle, there are $k+1$ possible values for $N_o$, but since
$\sum_{i=1}^k n_i=n_{is}$, the parity of $N_o$ is the parity of
$n_{is}$. Thus, for $k$ odd, there are $(k+1)/2$ possible values
of $I(n)$ for each value of $n_{is}$. For $k$ even there are
$\frac{k}{2}+1$ possible values for $I(n)$ when $n_{is}$ is even,
and $k/2$ values when it is odd. Remarkably, this is precisely the
number of possible values of $I(n)$ that are found in the lowest
order interference pattern for the Read-Rezayi states (see Eq.
(\ref{visibilityrr}) and the discussion around it).

As in the $k=2$ case, then, for all values of $k$ the number of
possible amplitudes for the interference pattern is the same for
the multi-component Halperin states and the Read-Rezayi states.
For $k\ge 3$ there is, however, a difference between the
amplitudes to be observed in the two states, as reflected in the
difference between Eqs. (\ref{visibilityrr}) and
(\ref{visibilityhalk}). And again, the results presented here for
the generalized Halperin state all depend on the symmetry between
the $k$ species of electrons, and thus lack the robustness of the
corresponding results for the Read-Rezayi states.

\section{Finite temperature Coulomb blockade}

In this section we study an interferometer in the Coulomb
blockaded limit, i.e., a quantum dot, and ask how the thermally
averaged number of electrons within the interferometer $\cal N$
and its derivative $\frac{\partial {\cal N}}{\partial A}(B,A,T)$
depend on the magnetic field, area and temperature. In the limit
$T=0$, the number of electrons on the dot is quantized to an
integer, and $\frac{\partial {\cal N}}{\partial A}(B,A,T)$ shows
Coulomb blockade peaks as a function of the area $A$. These peaks
may be smeared either by opening the point contacts that define
the dot, approaching the lowest order interference limit discussed
in previous sections, or by raising the temperature, as we discuss
now. In the analysis below we assume that the edge is fully
decoupled from the bulk, such that the state of the
quasi-particles in the bulk remains constant for a time long
enough for the measurement to take place, while the state of the
edge is in thermal equilibrium with its environment.

\begin{figure}[t]\includegraphics[width=1.2\linewidth]{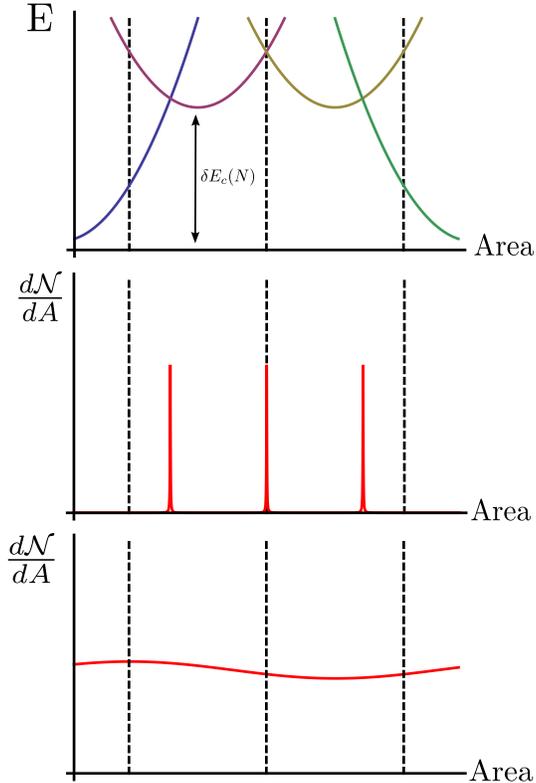}
\caption{The behavior of $\frac{\partial {\cal N}}{\partial
A}(B,A,T)$ as a function of area in three different temperature
regimes. In the low and intermediate temperature regimes the
electron number $N$ is well quantized. The upper graph plots the
energy parabolas, each of which corresponds to a different value
of $N$.  The contribution $\delta E_c$ of Eq. (\ref{deltaec}) is
indicated. Peaks in $\frac{\partial {\cal N}}{\partial A}(B,A,T)$
occur at crossing points of the parabolas. These points are
unevenly spaced due to $\delta E_c$ (middle graph, solid lines).
In the intermediate temperature regime, the peaks are broadened
and shift towards even spacing (the dashed lines indicate the
limit of even spacing).  In the high temperature regime (the lower
graph) the peaks are smeared to small oscillations. The plots use
the case $k=3$, $n_{is}=0$ as an illustration.}
\label{fig:FiniteT}
\end{figure}

We start by studying $\frac{\partial {\cal N}}{\partial A}(B,A,T)$
for multi-component Halperin states, and continue with the
Read-Rezayi states. For both types of states at zero temperature
the peaks are unevenly spaced. In the intermediate temperature
regime, where the temperature is lower than the dot's charging
energy but higher than the typical energy for the dot's neutral
modes, the peaks are well defined, yet they are broadened and
shift towards equal spacing. When the temperature increases
further to exceed the dot's charging energy, the peaks are
smeared, and $\frac{\partial {\cal N}}{\partial A}(B,A,T)$ shows
only small oscillations as a function of the area. This behavior
is illustrated in Fig. (\ref{fig:FiniteT}). As we show below, both
the deviation of the peaks from equal spacing in the intermediate
temperature range and the small oscillations in the high
temperature regime carry the same information on the state of the
system as the lowest order interference discussed in the previous
sections.

\subsection{Multi-component Halperin states}

Generally, the Hamiltonian density for an abelian multi-edge
quantum Hall state is of the multi Luttinger liquid form,
%
%***************** interaction multi-component  ****************
\begin{equation}
{\cal H}=\frac{1}{4\pi}\sum_{ij}V_{ij}\Big
(\partial_x\phi_i-\frac{\Phi_i}{L}\Big )\Big
(\partial_x\phi_j-\frac{\Phi_j}{L}\Big )
%{\cal L}= K_{ij}\partial_t\phi_i\partial_x\phi_j-V_{ij}\partial_x\phi_i\partial_x\phi_j
\label{hamdensity}
\end{equation}
%************************************************************************
%
where the equilibrium values $\Phi_i/L$ are determined by the
magnetic field and the area enclosed by the edge. Unless otherwise
defined, sums go over the range $1..k$. The fields $\phi_j$
satisfy the commutation relations
%
%*************** commutator multicomponent **************
\begin{equation}
\left [ \phi_i(x),\partial_x\phi_j(x')\right ]=2\pi
K^{-1}_{ij}\delta(x-x') \label{commrel}
\end{equation}
%*********************************************************
%
In the absence of bulk quasi-particles the boundary conditions of
the $\phi$ fields are
%
%*****************************  boundary condition without QPs  **************
\begin{equation}
\phi_i(x+L)=\phi_i(x)+2\pi m_i \label{pbc1}
\end{equation}
%***********************************************************
%
with $m_i$ integers. The symmetry between the different $k$
electron flavors, together with the Hall conductance being
$k/(k+2)$, requires the equilibrium values of $\partial_x\phi_i$,
to satisfy $\frac{\Phi_i}{L}=\frac{BA}{L(k+2)\phi_0}$, independent of
$i$. We will keep on using the notation $\Phi$, but omit the
subscript. Furthermore, the symmetry implies a high degree of
symmetry for $V_{ij}$. Here we take $V_{ij}=V_1+V_2\delta_{ij}$.
The energy corresponding to a set $\{ m_i\}$ of winding numbers is
%
%*********************  energy without QPs ******************
\begin{equation}
E(\{m_i\})=\frac{\pi  V_1}{L}\Big [\sum_i (m_i-\Phi)\Big
]^2+\frac{\pi V_2}{L}\sum_i(m_i-\Phi)^2 \label{lowestenergy}
\end{equation}
%********************************************************
%
There are $k$ electron creation operators whose scaling dimensions
are lowest. These operators are $e^{il^{(j)}K\phi}$, with
$j=1..k$. Each of these operators changes the winding number $m_j$
by one.

The bulk quasi-particles introduce a shift in the boundary
conditions of these fields, making them
%
%************** boundary condition with QPs ***************
\begin{equation}
\phi_i(x+L)=\phi_i(x)+2\pi m_i+2\pi l^{(i)}K^{-1}n \label{pbc2}
\end{equation}
%*****************************************************
%
with the vector $n$ describing the bulk quasi-particles. This
shift may be understood by noting that when the quasi-particles of
the vector $n$ are created in the bulk, the state of the edge is
affected by a creation operator that is a superposition of
operators of the type $ e^{in_i\phi_i(x)}$ for different positions
$x$'s. For every position $x$, this operators leads to the shift
(\ref{pbc2}). In the presence of that shift, the energies
(\ref{lowestenergy}) change to
%
%*************  energy with QPs ***********************
\begin{eqnarray}
E(\{m_i\})= & \frac{\pi V_1}{L}\left [\sum_i
(m_i-\Phi-l^{(i)}K^{-1}n)\right ]^2+\nonumber \\ & \frac{\pi
V_2}{L}\sum_i(m_i-\Phi-l^{(i)}K^{-1}n)^2
\label{lowestenergywithqp}
\end{eqnarray}
%********************************************************
%
As seen in Eq. (\ref{lowestenergywithqp}) the very same term
$l^{(i)}K^{-1}n$ that introduced a phase to the wave function of
the interfering quasi-particle in the limit of lowest order
interference appears now as a flux shifting the energy for adding
electrons to the edge in the limit of a closed dot.

With the spectrum at hand, we now analyze the thermodynamics of a
closed dot as a function of temperature and area. In particular,
we look at the way the average number of electrons ${\cal N}(A,T)$
on the dot depends on area $A$ and the temperature $T$. The
average $\cal N$ is calculated thermodynamically by summing over
all configurations $\{m_j\}$. For a configuration $\{m_j\}$ the
number of electrons on the edge of the dot is
$N(\{m_j\})\equiv\sum_j m_j$. It is useful to express $\cal N$ in
terms of the canonical-ensemble partition function $Z_N$, in which
the contributing configurations all have $N(\{m_j\})=N$:
\begin{equation}
Z_N\equiv{\sum_{\{m_j\}}\exp{-\frac{E({\{m_j\}})}{T}}}\
\delta_{N,N(\{m_j\})} \label{canens}
\end{equation}
The thermodynamical average $\cal N$ is then
\begin{equation}
{\cal N}=\frac{\sum_{N}NZ_N}{\sum_NZ_N}
\label{partritionfunctiondefinition}
\end{equation}

At zero temperature, $\cal N$ is the integer number of electrons
($\sum_i m_i$) that minimizes the energy
(\ref{lowestenergywithqp}). When this number changes, a Coulomb
blockade peak appears.

There are two energy scales that define the temperature regimes in
the problem. The lower one is the scale of neutral degrees of
freedom, $V_2/L$, and the higher one is the scale associated with
charging energy, $V_1/L$. For a temperature that is much lower
than both scales we can approximate $T\approx 0$.

In the intermediate regime $\frac{V_1}{L}\gg T\gg \frac{V_2}{L}$
the number of electrons on the dot is still approximately
quantized, but many configurations $\{m_i\}$ contribute. For an
area $A$ for which the number of electrons on the dot is quantized
to a value $N_0$ the partition function of the dot $Z_{N_0}$ may
be calculated in the canonical ensemble, and involves a summation
over all internal states of the dot under the constraint that its
total number of electrons is $N_0$. For an area $A$ for which the
dot is close to a transition from ${\cal N}=N_0$ to ${\cal
N}=N_0+1$ the sum in (\ref{partritionfunctiondefinition}) includes
only the terms $N_0$ and $N_0+1$ and we have
%
%*********** particle number general *************
\begin{equation}
{\cal N}=N_0+\frac{Z_{N_0+1}}{Z_{N_0}+Z_{N_0+1}}
\label{neutralshift}
\end{equation}
%******************************************
%
We calculate $Z_N$ in the Appendix. We find it convenient to write
it in a form that highlights the contribution of the neutral mode.
We find
%
%********** partition function neutral energy  ****************
\begin{equation}
Z_N\approx {2 \pi \over \sqrt{k}} \Big({T L \over V_2}\Big)^{k-1
\over 2}\exp{-\frac{E_c(N) +\delta E_c(N)}{T}} \label{ZN}
\end{equation}
%*******************************************************
%
In (\ref{ZN}) the pre-factor does not depend on $N$, and hence
does not affect our calculation of $\cal N$. This prefactor
originates from the entropy associated with the different
configurations of the neutral mode. Within the exponential factor,
$E_c$ is the charging energy in the absence of a neutral mode
%
%***********************  charging energy  *********************
\begin{equation}
E_c(N)={\pi \over  L} \big(V_1 + {V_2\over k}\big) \Big [N -
\sum_j \big({1\over k + 2} {B A \over \Phi_0} \ + \ l^{(j)} K^{-1}
\; n\big) \Big]^2 \label{ec}
\end{equation}
%**************************************************************
%
and  $\delta E_c$ is
%
%************************************ shift of free energy  due to neutral excitations **********
\begin{equation}
\delta E_c=- 2 T k I(n)  e^{-\frac{\pi TL}{V_2}\left( 1 - {1 \over
k}\right) }\
%{\mbox Re}\left [e^{i\frac{2\pi N}{k}}\ \sum_j e^{i2\pi l_jK^{-1}n}\right ]
\cos \Big( 2 \pi {N \over k} - \pi {n_{is} \over k} \Big)
\label{deltaec}
\end{equation}
%***************************************************************************************************
%
The factor $e^{-\frac{\delta E_c}{T}}$ originates from the $\cal
N$-dependent energy and entropy that come out of the different
configurations of the neutral mode associated with the same
electron number.

As evident from Eqs. (\ref{neutralshift}) and (\ref{ZN}), the
center of the peak in $\frac{\partial {\cal N}}{\partial A}$ is at
the area for which $E_c(N) +\delta E_c(N)=E_c(N+1) +\delta
E_c(N+1)$ (see upper graph of Fig.~(\ref{fig:FiniteT})). As seen
in (\ref{deltaec}), the neutral mode contribution $\delta E_c$,
which is of the order of $V_2/L$ at zero temperature, becomes
exponentially small at the intermediate temperature regime, and
the peaks in $\frac{\partial {\cal N}}{\partial A}$ approach equal
spacing, with the correction to equal spacing being exponentially
small in $TL/V_2$. The correction oscillates with the number of
electrons on the dot, with a period of $k$ electrons, and depends
on the state of the bulk quasi-particles through the factor
%$\sum_j e^{i2\pi l^{(j)}K^{-1}n}$,
$I(n)$, the same factor that appears in the lowest order
interference term, see Eq. (\ref{visibilityhalk}).
%(\ref{statisticsk}).

When the temperature is higher than the charging energy, $\cal N$
is not quantized. The thermally averaged $\cal N$ may be
calculated through the standard methods to be
%
%************************  particle number high temperature limit ******************
\begin{eqnarray}
{\cal N}  & =&  \frac{k}{k+2}\frac{BA}{\Phi_0}+{\cal
N}_1\Big(\frac{BA}{\Phi_0}\Big) \label{highT} - {2  T L I(n) \over
V_1 + V_2/k }
\\[.5cm]
& & %\hspace*{-1.5cm}
\times e^{- {\pi T L \over  k^2 (V_1 + {V_2 \over k})} - {\pi T
L(1 - {1 \over k} )  \over  V_2}} \sin\Big( \frac{2\pi
BA}{(k+2)\Phi_0}-\frac{\pi n_{is}}{k+2}\Big)\nonumber
\end{eqnarray}
%***************************************************************************************
%
with $I(n)$ the interference visibility, whose definition and
properties are given at and below Eq. (\ref{visibilityhalk}). In
this expression the first term is the uniform linear increase of
the charge with area, for a fixed density. The second and third
terms are corrections that fall off exponentially in the limit of
high temperature. The second term  results from discreteness of
the charge, but is insensitive to the neutral modes (i.e., it is
independent of $V_2$). The period of this term is an area increase
that corresponds to a single electron. The third term is the one
that we are interested in. The quasi-particle properties reflected
in this term are precisely those that are reflected by the
interference phase, Eq. (\ref{statisticsk}). This limit is
illustrated in the bottom graph of Fig.~(\ref{fig:FiniteT}).

\subsection{Read-Rezayi states}

A similar calculation may be carried out for the non-abelian
Read-Rezayi states. Explicit expressions for the partition
function are cumbersome. However, their low temperature and high
temperature limits are rather easy to calculate. In fact, the two
limits are related by a remarkable set of identities, derived by
Cappelli et al. in (\onlinecite{Cappelli-2}), based on the modular
invariance of the partition functions of the edge theory of the
Read-Rezayi states.

The partition function for an edge of a Read-Rezayi state depends
on the state of the bulk, and is characterized by two quantum
numbers, $n_{is},l$. Neglecting normalization factors that do not
depend on these quantum numbers\cite{Cappelli-2},
%
%*****************  partition function RR *******************************
\begin{eqnarray}
Z_{n_{is}}^l & =& \sum_{p=-\infty}^\infty\sum_{b=1}^{k} e^{-
\frac{\pi V_1(k+2)}{TL}\frac{\left
[pk+b-\frac{n_{is}}{k+2}-\frac{k}{k+2}\frac{BA}{\Phi_0}\right
]^2}{ k}} \nonumber
\\[.2cm]
& &  \times \chi_{n_{is}+2b}^l(T) \label{partitionfunction}
\end{eqnarray}
%*************************************************
%
This expression is an analog to Eq.
(\ref{partritionfunctiondefinition}). However, while for the
multi-component states we had to sum over $k$ quantum numbers, one
per each edge state, for the Read-Rezayi states there are only two
edge modes, one charged and one neutral. The two sums in
(\ref{partitionfunction}) combine to a sum over all possible
charges on the charged mode.  The parafermionic character of the
neutral part $\chi_m^l$ includes a sum over all internal states of
the neutral mode. It is periodic as a function of $m$, with a
period of $2k$.The parameter $l$ is an integer in the range $0\le
l\le k$, and $n_{is}+l$ is an even number.

In the absence of the $\chi$ factor, $Z_{n_{is}}^l$ is a partition
function of a single chiral Luttinger liquid, where the number of
electrons on the dot is $kp+b$, the number of quasi-holes in the
bulk is $n_{is}$, and the charge of a quasi-hole is $1/(k+2)$. The
character of the neutral part depends on two quantum numbers. One
of them, $l$, depends only on the state of the bulk, while the
other depends also on the number of electrons in the dot, as
reflected in the factor $n_{is}+2b$.

In the limit of $T\rightarrow 0$ the partition function is
relatively easy to calculate. The contribution of the neutral part
of the edge is
%********************  parafermion character *********************
\begin{equation}
\chi_m^l(T)=\exp{\left ({-\frac{\epsilon_{gs}^{m,l}}{T}}\right )}
\label{teqzeroneutral}
\end{equation}
%**************************************
%
where $\epsilon_{gs}^{m,l}$ is the ground state energy of the edge
in the topological sector defined by $m,l$. The ground state
energy is dictated by the conformal field theory that describes
the edges of Read-Rezayi states, $Z_k$ parafermions. It is
determined by the conformal dimension $h^l_m$ of the parafermionic
field that corresponds to the topological sector $m, l$ to be,
%********************  parafermion character *********************
\begin{equation}
\epsilon_{gs}^{m,l}=\frac{2\pi V_2}{L}h_m^l=\frac{2\pi
V_2}{L}\Big(\frac{l(l+2)}{4(k+2)}-\frac{m^2}{4k}\Big)
\label{teqzeroneutral2}
\end{equation}
%**************************************
%

As emphasized by Cappelli et al. in Refs.
[\onlinecite{Cappelli-1,Cappelli-2}], due to the modular
invariance of the partition function, the parafermionic characters
at high temperature are related to those at low temperature
through the modular S-matrix. Generally, the characters are a
function of the dimensionless variable $\tau=V_2/LT$. Their
modular invariance dictates,
\begin{equation}
\chi_m^l(\tau)=\sum_{l^\prime=0}^k\sum_{m^\prime=-k}^{k-1}S_{m,l;m',l'}\chi_{m'}^{l'}(1/\tau)
\label{modinvgeneral}
\end{equation}
where $S_{m,l;m',l'}$ are elements of the modular $S$ matrix,
\begin{equation}
S_{m,l;m',l'}=\frac{1}{\sqrt{k(k+2)}}\sin{\pi\frac{(l+1)(l'+1)}{k+2}}e^{-i\pi\frac{mm'}{k}}
\label{smatrix}
\end{equation}

The high temperature limit $T\gg\frac{V_2}{L}$ of $\chi_m^l$ is
then obtained by substituting Eq. (\ref{teqzeroneutral}) into
(\ref{modinvgeneral}) to get
%
%********** modular invariance  *****************
\begin{equation}
\chi_m^l=\sum_{l^\prime=0}^k\sum_{m^\prime=-k}^{k-1}S_{m,l;m',l'}e^{-\frac{2
\pi TL}{ V_2}h_{m^\prime}^{l^\prime}} \label{modularinvariance}
\end{equation}
%******************************************
%

The leading order contributions in the limit of $T\gg
\frac{V_2}{L}$  come from the terms $l'=m'=0$  and $l'=m'=1$ and $l'=-m'=1$.
For the former, which we will call "the identity ($\openone$)
term", $h_0^0=0$. For the latter, which we will call "the
quasi-particle term", $h_{\pm 1}^1=\frac{k-1}{2k(k+2)}$. Limiting
ourselves to these two terms and substituting Eqs.
(\ref{modularinvariance}) and (\ref{smatrix}) in
(\ref{partitionfunction}) we can calculate the high temperature
expansion of $Z_{n_{is}}^l$. We find $Z=Z_{\openone} + Z_{qp}$
with the two terms corresponding to the identity and
quasi-particle contributions, respectively. Since $Z_{\openone}\gg
Z_{qp}$, we have
$\ln{Z}=\ln{Z_{\openone}}+\frac{Z_{qp}}{Z_{\openone}}$. Again, we
extract the dependence of $\cal N$ on $A$ by taking the derivative
$\frac{\partial \ln Z}{\partial A}$. Equation
(\ref{modularinvariance}) limits us to $T\gg V_2/L$, but allows us
to calculate both the intermediate ($T\ll V_1/L$) and high ($T\gg
V_1/L$) temperature regimes.

In the intermediate temperature regime Eqs. (\ref{neutralshift})
and (\ref{ZN}) hold just as they did for the multi-component
Halperin states, but $\delta E_c$ should be re-calculated. Coulomb
blockade peaks are still pronounced, although they are broadened
and shifted. Now, the shift in the charging energy corresponding
to the dot having $N$ electrons depends on the quantum number $l$
of the bulk quasi-particles. It is,
%
%**************  free energy shift RR  ****************
\begin{equation}
\delta E_c=4T e^{-\frac{2 \pi TLh_1^1}{V_2}}\
\cos{\Big({\frac{2\pi
N}{k}-\pi\frac{n_{is}}{k}}\Big)}\cos{\Big(\pi\frac{l+1}{k+2}\Big)}
\end{equation}
%************************************
%
The suppression factor of the interference, Eq.
(\ref{visibilityrr}) appears here as determining the shift of the
Coulomb blockade peak.

In the high temperature regime the result is similar to Eq.
(\ref{highT}). To leading order ${\cal
N}\approx\nu\frac{BA}{\Phi_0}$. The first correction ${\cal N}_1$
has a periodicity that corresponds to the addition of one
electron, and is independent of $V_2/L$. The second correction is
the one we are interested in. It is,
%
%**********************  neutral correction to particle number RR ********
\begin{eqnarray}
- {2 T L \over V_1(k+2)} \cos \Big(\pi {l+1 \over k+2} \Big) e^{-
{2 \pi T L h_1^1 \over V_2}
- { \pi T L \over V_1 k (k+2)}}\nonumber \\[.5cm]
\times \sin\Big( \frac{2 \pi}{k+2} \frac{BA}{\phi_0}
-\pi\frac{n_{is}}{k+2}  \Big) \label{rrhight}
\end{eqnarray}
%*****************************************************************
%
Again, the dependence of this term on the state of the environment
is identical to that found for the lowest order interference.
Furthermore, the thermal suppression factor in (\ref{rrhight}) is
identical to that found in the limit of lowest order interference
(see [\onlinecite{BisharaNayak}] for the $k=2$ case). The high
temperature remainder of the quantization of the charge in the
quantum dot is thus found to be intimately connected to the lowest
order interference.

\section{Summary}

For non-interacting electrons at $\nu=1$, the transition from
lowest order interference in an open Fabry-Perot interferometer to
a discrete spectrum in a closed one may be understood by
Bohr-Sommerfeld semi-classical arguments. These arguments make it
possible to describe the formation of the discrete state through
the interference of infinitely many trajectories. This line of
thought cannot be immediately applied when interactions are
involved and the interferometer is in a fractional quantum Hall
state, with quasi-particles that carry a fractional charge. This
reasoning is hard to apply for complicated fractional quantum Hall
states, where many edge modes co-exist. In all these cases,
several types of quasi-particles may tunnel across the
constriction, but the resonances that form when the interferometer
closes to be a quantum dot are resonances that correspond to
adding or removing electrons.

The behavior of the number of electrons in the interference loop
$\cal N$ as the area, magnetic field and temperature are varied,
allows us to explore the universal aspects of the transition, as a
function of temperature, from sharp resonances to sinusoidal
behavior in an interferometer in the Coulomb blockaded limit. This
is possible since in that limit $\cal N$ does not depend on the
non-universal aspects of the two constrictions that form the
interferometer, such as the matrix elements for the tunneling of
different types of quasi-particles. Rather, it is determined by
the partition function, whose modular invariance exposes its
universal high temperature properties. As Eqs.
(\ref{teqzeroneutral},\ref{modularinvariance}) show, in the
Read-Rezayi states the high temperature partition function is
composed of several leading terms, with each one corresponding to
one revolution of the interference loop by one type of
quasi-particle, exactly like the lowest order interference term in
an open interferometer. The relative weight between the different
quasi-particles is determined by the elements of the modular
$S$-matrix and by the thermal suppression of the various terms.
The latter, in turn, is determined by the conformal dimensions of
the corresponding quasi-particles. The high temperature partition
function of the multi-component Halperin states, too, is a sum of
such terms (see the Appendix below), that can be mapped onto
interference of the various types of quasi-particles, each winding
once around the interferometer. For both types of states, the
thermally smeared Coulomb blockade peaks carry the same
information as the lowest order interference about the topological
properties of the state.

The comparison of the interference and Coulomb blockade patterns
predicted for the abelian multi-component Halperin states and the
non-abelian Read-Rezayi states show that when the former are at a
symmetry point between the $k$ electronic flavors that compose
them, the interferometer cannot distinguish them from the latter.
This observation highlights the crucial difference between the two
types of states. The properties of the Fabry-Perot interferometer
for the non-abelian states are insensitive to local perturbations
to its Hamiltonian, while the properties of the abelian states are
modified when local perturbations to its Hamiltonian shift it away
from the symmetry point.

Finally, we comment on common experimental values for the
parameters we use. The calculations we carry out are valid at
temperatures much smaller than the bulk energy gap. For the
$\nu=5/2$ state this gap is around $0.5$K. The two velocities
$V_1$ and $V_2$ were recently calculated numerically by Hu {\it
{et al.}}\cite{PffafianNumerics}, who found $V_2\approx
10^6$cm/sec, and $V_1/V_2\approx 7$ for a quantum dot at
$\nu=5/2$. The two energy scales $V_2/L$ and $V_1/L$ should be
both smaller than the gap, and yet large enough to be within reach
of electron cooling. These requirements constrain the dot to be of
a circumference of several microns, a scale that is within
experimental reach.

\section{Acknowledgments}

We are grateful to Andrea Cappelli for an instructive discussion
and for sharing with us unpublished notes. This work was supported
by the US-Israel Binational Science Foundation, by the Minerva
foundation, by the Einstein center at the Weizmann institute, by
Microsoft's Station Q, by NSF grant DMR-0906475 and by the
Heisenberg program of the DFG.

\vskip 1cm

\appendix

\section{Partition function for multi-component Halperin states}

\subsection{Canonical ensemble}

In this subsection we calculate the partition function $Z_N$ for a
quantum dot in a multi-component Halperin state in the canonical
ensemble, in which the dot has $N$ electrons.

We introduce the abbreviations
%
%*************************  background charge  **********************
\begin{eqnarray}
s_{j,0} & \equiv &  {1\over k + 2} {B A \over \Phi_0} \ + \ l^{(j)}\; K^{-1} \; n \\
& = &  {1\over k + 2} {B A \over \Phi_0}         - {n_{is} \over
2(k+2)} \ + \ {1 \over 2} n_{j}        \ \ . \nonumber
\end{eqnarray}
%********************************************************************
%
Here, $n_{j}$ denotes the number of QPs of type $j$ inside the
droplet, and $n_{is} = \sum_j n_{j}$ is the total number of QPs.
We note that
%
%************************  some over background charge  *************
\begin{equation}
S_0 \ \equiv\  \sum_j s_{j,0} \ = \   {k\over k + 2} {B A \over
\Phi_0}    \ + \ n_{is} \ {1 \over k+2} \ \ . \label{S0.eq}
\end{equation}
%***********************************************************************
%

We calculate the canonical partition function by enforcing the
constraint of fixed particle number $N$ by a Lagrange multiplier.
Then, we perform an unconstrained sum over the set of winding
numbers $\{m_i\}$ by applying the Poisson summation formula to
each of the sums over winding numbers.
%
%*************** partition function *****************
\begin{widetext}
\begin{eqnarray}
Z_N & = & e^{- {\pi V_1\over T L} (N - \sum_j s_{j,0})^2}\ \int d
\lambda \sum_{\{s_j\}} e^{- {\pi V_2 \over T L}
\sum_j (s_j - s_{j,0})^2   + i \lambda N - i \lambda \sum_j s_{j}} \\[.5cm]
& = &{2 \pi\over \sqrt{k}} \Big({T L\over  V_2}\Big)^{k-1 \over 2} e^{- {\pi \over T L} (V_1 + {V_2\over k}) (N - \sum_j s_{j,0})^2}\ \nonumber \\[.5cm]
& & \hspace*{2cm} \times \sum_{\{p_i\}}  \exp \Big[ { \pi T L
\over V_2} \big( {1 \over k} (\sum_j p_j)^2 - \sum_j p_j^2 \big) +
{2 \pi i \over k} \sum_j p_j (N - \sum_l s_{l,0}) + 2\pi i  \sum_j
p_j s_{j,0} \Big] \nonumber
\end{eqnarray}
\end{widetext}
%**********************************************
%
If $T L/V_2  \gg 1$, leading terms in the partition function have
all $p_j\equiv \overline{p}$ equal to each other. The partition
function depends on $\overline{p}$ only via a term $\exp(2 \pi i
\overline{p} N) \equiv 1$. Summing over $\overline{p}$ thus
introduces an infinite normalization factor. The origin of this
factor is the fact that we introduced $k$ Poisson variables
$\{p_j\}$ while only $k-1$ edge charges are freely summed over due
to the constraint of fixed total particle number. Having
understood the mathematical reason for the appearance of this
infinite normalization factor, we discard it and consider
$\overline{p}=0$ in the following. We denote the leading term with
all $p_j =0$ by
%
%****************  leading part partition function *******************
\begin{equation}
Z_N^{(0)} \ = \ {2 \pi \over \sqrt{k}} \Big({T L \over
V_2}\Big)^{k-1 \over 2} \ e^{- {\pi \over T L} (V_1 + {V_2\over
k}) (N - \sum_j s_{j,0})^2} \ \ . \label{partzero.eq}
\end{equation}
%*******************************************************************
%

Next, we turn to the subleading terms. The  most important of
these terms have $p_j = \pm \delta_{j,j_0}$ with $j_0 =
1,2,...,k$, i.e.~one of the $\{p_j\}$ different from zero. Noting
that
%
%********************** QP sum modulo 2 **************
\begin{equation}
{1 \over k} \sum_j \cos(\pi n_{j})  = I(n) \ \ ,
\end{equation}
%*******************************************************
%
this contribution reads
%
%*************** subleading partition function **************************
\begin{eqnarray}
Z_N^{(1)}&   = & {2 \pi\over \sqrt{k}} \Big({T L \over
V_2}\Big)^{k -1 \over 2}  \ e^{- {\pi \over T L} (V_1 + {V_2\over
k}) (N - \sum_j s_{j,0})^2}
\label{partone.eq} \\
& & \times e^{- {\pi T L \over  V_2} (1 - {1 \over k})} \ 2 k I(n)
\cos\Big( 2 \pi {N \over k} - \pi {n_{is} \over k} \Big) \ \ .
\nonumber \label{subleading}
\end{eqnarray}
%***************************************************
%

\subsection{Periodicity at intermediate temperatures }

 We now use the expression Eq.~(\ref{neutralshift}) to determine
the location of the Coulomb blockade (CB) peak from the
requirement $\langle N \rangle = N_0 + {1\over 2}$, which
corresponds to demanding that the ratio of partition functions
$Z_{N_0+1}/Z_{N_0}=1$ is unity. This condition determines the area
corresponding to a CB peak as
%
%********************  area of CB peak  *****************
\begin{eqnarray}
A(N_0)  & = & {k+2 \over k} {\Phi_0 \over B} \left\{N_0 - {n_{is}
\over k + 2}  + {1 \over 2} + {I(n) T L \over \pi  (k V_1 +  V_2
)}
\right. \nonumber \\[.5cm]
%& &
%\  \nonumber
%\\
& &\hspace*{-2.2cm}\left. \times e^{- {\pi T L (k-1)\over k V_2} }
\Big[ \cos\big(2 \pi {N_0 +1 - { n_{is} \over 2}  \over k} \big)
- \cos\big( 2 \pi {N_0   -  {n_{is}\over 2} \over k} \big)\Big] \right\} . \nonumber\\[.5cm]
\label{peaklocation.eq}
\end{eqnarray}
%**********************************************************
%
When calculating  the Fourier transform of a sequence of $k$
consecutive CB peaks with locations according to
Eq.~(\ref{peaklocation.eq}), the leading harmonic is determined by
the term proportional to $I(n)$ and thus has the same suppression
factor as the lowest order interference.

\subsection{Grand canonical ensemble}

In the high temperature limit the number of electrons on the dot
fluctuates thermally. We now use the results
Eqs.~(\ref{partzero.eq}), (\ref{partone.eq}) to calculate the
grand canonical partition function for high temperatures $T \gg
V_1/L, V_2/L$. There will be two contributions with a nontrivial
area dependence: using the Poisson summation formula to  sum
$Z_0(N)$ over particle number, the contributions with Poisson
index  $ n = \pm 1$ describe the high temperature limit of
standard CB with a  periodicity of one electron. The contribution
$Z_1(N)$ on the other hand is already exponentially small in
$T/V_2$, hence it can be integrated over the number of particles
and there is no need to use the Poisson summation formula for this
term.  The area dependence of Eq. (\ref{subleading}) has  a period
of $k$ electrons. We first calculate the grand canonical
generalization of the partition function Eq.~(\ref{partzero.eq}).
To simplify notation, we use the abbreviation $S_0$ introduced in
Eq.~(\ref{S0.eq}).
%
%********************* grand canonical Z_0 *********************************
\begin{eqnarray}
Z_0(\mu) & = & {2 \pi \over \sqrt{k}} \Big({T L \over
V_2}\Big)^{k-1 \over 2} \ \sum_N\  e^{- {\pi \over L T}(V_1 + {V_2
\over k})(N - S_0)^2
+ \frac{\mu N}{T}}\nonumber \\[.5cm]
&\hspace*{-1cm}  \approx & \hspace{-0.5cm}   { 2 \pi \over
\sqrt{k}} \Big({T L  \over  V_2}\Big)^{k-1 \over 2} \sqrt{T L
\over V_1 + {V_2 \over k}}\ \
e^{ { \mu^2  L \over 4 \pi T (V_1 + {V_2 \over k})}  \ +\  {\mu S_0\over T}} \nonumber \\[.5cm]
& &  \hspace*{-1cm}   \times
\left\{1 +  e^{-{\pi T L \over V_1 + {V_2 \over k}}} \ 2 \cos\Big[2 \pi\big(S_0 + {\mu L \over 2 \pi ( V_1 + {V_2 \over k})}\big)\Big]\right\} \nonumber\\[.5cm]
\end{eqnarray}
%*******************************************************************************
%
We next calculate the grand canonical generalization of the
partition function Eq.~(\ref{partone.eq})
%
%**********************  grand canonical part one **************************
\begin{widetext}
\begin{eqnarray}
Z_1(\mu) & = & {4 \pi \sqrt{k}}  I(n) \Big({T L  \over
V_2}\Big)^{k-1 \over 2} \ e^{- {\pi T L  \over  V_2} (1 - {1 \over
k})} \int dN e^{- {\pi \over L T} (V_1 + {V_2 \over k})(N - S_0)^2
+ { \mu N\over T}}  \cos\Big( 2 \pi {N \over k} - \pi {n_{is} \over k} \Big) \nonumber \\[.5cm]
& = & {4 \pi  \sqrt{k}} I(n)  \Big({T L \over  V_2}\Big)^{k-1
\over 2} \ \sqrt{T L  \over V_1 + {V_2 \over k}}\ \ e^{ { \mu^2 L
\over 4 \pi T (V_1 + {V_2 \over k})}  \ +\  { \mu S_0\over T}}\
e^{- {\pi T L \over  V_2} (1 - {1 \over k})} \nonumber \\[.5cm]
& & \hspace{4cm} \times e^{- {\pi T L  \over  k^2 (V_1 + {V_2
\over k})}} \cos\Big[{2 \pi\over k}\big(S_0 + { \mu L  \over 2 \pi
( V_1 + {V_2 \over k})}- {n_{is}\over 2}\big)\Big]
\end{eqnarray}
\end{widetext}
%******************************************************************************
%
Combining the two parts, we find (up to a constant) for the
logarithm of the partition function
%
%**************************** logarithm of partition function **************
%\begin{widetext}
\begin{eqnarray}
\ln Z(\mu) & = &  {\mu^2 L  \over 4\pi T( V_1 + {V_2 \over k})} \
+ \ { \mu S_0 \over T} \label{logpart.eq}
\\[.5cm]
& + & e^{- {\pi T L  \over V_1 + {V_2 \over k}}} \ 2 \cos\Big[ 2
\pi \big( S_0 + { \mu L \over 2\pi (V_1 + {V_2 \over k})}\big)
\Big] \nonumber
\\[.5cm]
& & + e^{- {\pi T L  \over k^2 (V_1 + {V_2 \over k})} - {\pi T L
\over V_2} (1 - {1 \over k})}
\  2 k I(n)  \nonumber \\[.5cm]
& & \times \cos\left[ {2 \pi \over k}\left(S_0 + {\mu  L \over 2
\pi (V_1 + {V_2 \over k})}  - {n_{is}\over 2} \right)\right]
\nonumber
\end{eqnarray}
%\end{widetext}
%*****************************************************************************
%
The particle number can now be calculated as a derivative ${\cal
N} = T {\partial \over \partial \mu} \ln Z |_{\mu =0}$.

%\bibliography{fp-cb}
\end{document}